\documentclass[aps,prb,twocolumn,superscriptaddress]{revtex4-2}

\usepackage{graphicx, color, xcolor}
\usepackage{amsmath,amsfonts,amssymb}
\usepackage{booktabs} %
\usepackage{soul}
\usepackage{hyperref}
\hypersetup{
    colorlinks=true,
    linkcolor=blue,
    filecolor=magenta,      
    urlcolor=cyan,
    pdftitle={QubitsNTD},
    }
\usepackage[heightadjust=all]{floatrow}
\usepackage{lineno}

\newcommand{\beginsupplement}{\setcounter{table}{0}  \renewcommand{\thetable}{S\arabic{table}} \renewcommand{\theHtable}{S\arabic{table}} \setcounter{figure}{0} \renewcommand{\thefigure}{S\arabic{figure}} \renewcommand{\theHfigure}{S\arabic{figure}}}

\providecommand{\ignore}[1]{}

\newcommand{\ket}[1]{|#1\rangle}

\newcommand{\baseline}{\ensuremath{100\,\text{points}}}
\newcommand{\effone}{$\epsilon^{selection}_{Q1}=(96 \pm 1)\,\%$}
\newcommand{\efftwo}{$\epsilon^{selection}_{Q2}=(92 \pm 1)\,\%$}

\def\lsim{\mathrel{\rlap{\lower4pt\hbox{\hskip1pt$\sim$}}
    \raise1pt\hbox{$<$}}}                
\def\gsim{\mathrel{\rlap{\lower4pt\hbox{\hskip1pt$\sim$}}
    \raise1pt\hbox{$>$}}}                

\begin{document}
\title{Quantitative characterization of superconducting qubits as particle detectors}
\author{Francesco De Dominicis}
\email{francesco.dedominicis@roma1.infn.it}
\affiliation{INFN -- Sezione di Roma, Roma I-00185, Italy}

\author{Raja Yasir Mehmood Khan}
\affiliation{Gran Sasso Science Institute, L’Aquila I-67100, Italy}
\affiliation{INFN – Laboratori Nazionali del Gran Sasso, Assergi (L’Aquila) I-67100, Italy}
\affiliation{Dipartimento di Fisica e Astronomia ``Galileo Galilei", Università degli Studi di Padova, Padova I-35131, Italy}

\author{Dounia L. Helis}
\affiliation{INFN – Laboratori Nazionali del Gran Sasso, Assergi (L’Aquila) I-67100, Italy}

\author{Ambra Mariani}
\affiliation{INFN -- Sezione di Roma, Roma I-00185, Italy}
\affiliation{ENEA Istituto Nazionale di Metrologia delle Radiazioni Ionizzanti (INMRI), Centro Ricerche Casaccia, I-00123, Rome, Italy}

\author{Letizia Tirabasso}
\affiliation{INFN -- Sezione di Roma, Roma I-00185, Italy}
\affiliation{Dipartimento di Fisica, Sapienza Universit\`a di Roma, Roma I-00185, Italy}

\author{Alberto Ressa}
\email{alberto.ressa@roma1.infn.it}
\affiliation{INFN -- Sezione di Roma, Roma I-00185, Italy}

\author{Mustafa Bal}
\affiliation{Superconducting Quantum Materials and Systems Division, Fermi National Accelerator Laboratory (FNAL), Batavia, IL 60510, USA}

\author{Fabio Bellini}
\affiliation{INFN -- Sezione di Roma, Roma I-00185, Italy}
\affiliation{Dipartimento di Fisica, Sapienza Universit\`a di Roma, Roma I-00185, Italy}

\author{Camilla Bonomo}
\affiliation{INFN -- Sezione di Roma, Roma I-00185, Italy}
\affiliation{Dipartimento di Fisica, Sapienza Universit\`a di Roma, Roma I-00185, Italy}

\author{Nicola Casali}
\affiliation{INFN -- Sezione di Roma, Roma I-00185, Italy}

\author{Gianluigi Catelani}
\affiliation{Institute for Theoretical Nanoelectronics (PGI-2), Forschungszentrum J\"ulich, 52428 J\"ulich, Germany}
\affiliation{Quantum Research Center, Technology Innovation Institute, Abu Dhabi 9639, United Arab Emirates}

\author{Ivan Colantoni}
\affiliation{Consiglio Nazionale delle Ricerche, Istituto di Nanotecnologia}
\affiliation{INFN -- Sezione di Roma, Roma I-00185, Italy}

\author{Francesco Crisa}
\affiliation{Illinois Institute of Technology}

\author{Angelo Cruciani}
\affiliation{INFN -- Sezione di Roma, Roma I-00185, Italy}

\author{Sabrina Garattoni}
\affiliation{Superconducting Quantum Materials and Systems Division, Fermi National Accelerator Laboratory (FNAL), Batavia, IL 60510, USA}

\author{Luca Gironi}
\affiliation{Universit\`a degli Studi di Milano-Bicocca, Dipartimento di Fisica, Milano, Italy}
\affiliation{INFN Sezione di Milano-Bicocca, Milano, Italy}

\author{Andrea Melchiorre}
\affiliation{Dipartimento di Scienze Fisiche e Chimiche, Universit\`a degli Studi dell'Aquila, I-67100 L'Aquila, Italy}
\affiliation{INFN – Laboratori Nazionali del Gran Sasso, Assergi (L’Aquila) I-67100, Italy}

\author{Lorenzo Pagnanini}
\affiliation{Gran Sasso Science Institute, L’Aquila I-67100, Italy}
\affiliation{INFN – Laboratori Nazionali del Gran Sasso, Assergi (L’Aquila) I-67100, Italy}

\author{Valerio Pettinacci}
\affiliation{INFN -- Sezione di Roma, Roma I-00185, Italy}

\author{Stefano Pirro}
\affiliation{INFN – Laboratori Nazionali del Gran Sasso, Assergi (L’Aquila) I-67100, Italy}

\author{Andrei Puiu}
\affiliation{INFN – Laboratori Nazionali del Gran Sasso, Assergi (L’Aquila) I-67100, Italy}

\author{Tanay Roy}
\affiliation{Superconducting Quantum Materials and Systems Division, Fermi National Accelerator Laboratory (FNAL), Batavia, IL 60510, USA}

\author{Shaojiang Zhu}
\affiliation{Superconducting Quantum Materials and Systems Division, Fermi National Accelerator Laboratory (FNAL), Batavia, IL 60510, USA}

\author{Anna Grassellino}
\affiliation{Superconducting Quantum Materials and Systems Division, Fermi National Accelerator Laboratory (FNAL), Batavia, IL 60510, USA}

\author{Laura Cardani}
\affiliation{INFN -- Sezione di Roma, Roma I-00185, Italy}

\date{\today}

\begin{abstract}
\textbf{Abstract:} 
Ionizing radiation is a major source of correlated errors in superconducting quantum processors, yet the mechanisms responsible for the qubit response to particle interactions remain only partially understood. 
In this work, we operate superconducting qubits as particle detectors while simultaneously monitoring the deposited energy with an independent semiconductor cryogenic sensor. This complementary measurement provides an absolute determination of the particle interaction rate in the chip, enabling the first direct measurement of the detection efficiency of superconducting qubits exposed to environmental radiation. We measure individual qubit detection efficiencies of about 30--40$\%$, increasing to approximately 50$\%$ when combining the response of two qubits. Under the conservative assumption that the efficiency loss is entirely determined by a finite detection threshold, we derive an upper limit on the effective energy threshold of the detector of $70\pm20~\mathrm{(stat)}\pm30~\mathrm{(syst)}~\mathrm{keV}$. Measurements with radioactive sources producing different deposited-energy spectra further show that the detection efficiency decreases for lower-energy interactions.
We further investigate the origin of the qubit response by injecting controlled thermal pulses with a resistive heater. Although these pulses deposit energies comparable to those released by particle interactions, they do not reproduce the radiation-induced signatures, demonstrating that the qubit response to radiation cannot be explained by a transient increase in substrate temperature alone. 
Our results establish an experimental framework for quantitatively connecting particle energy deposition to radiation-induced responses in superconducting quantum circuits.
\end{abstract}

\maketitle
\section{Introduction}
Superconducting quantum circuits are among the leading platforms for scalable quantum information processing, with rapid progress demonstrated in multi-qubit processors and error-corrected architectures~\cite{Barends2014, Arute:2019, Kjaergaard2020, RigettiAspen, wu:2021, IBM:eagle}. 
As coherence times improve and devices scale up, previously subdominant noise sources are becoming increasingly relevant. In particular, the interaction of superconducting qubit chips with particles originating from environmental radioactivity and cosmic rays~\cite{demetra:2020} has attracted growing attention.
Exposure to elevated radiation levels has been shown to reduce the lifetime of transmon qubits~\cite{Vepsalainen:2020}. Energetic particle interactions in the substrate generate bursts of high-energy phonons, which propagate across the chip and break Cooper pairs, producing nonequilibrium quasiparticles that induce correlated errors in multiple qubits~\cite{McEwen:2021, Wilen:2021}. 
More recently, radiation-induced events have also been shown to produce correlated frequency shifts lasting up to the millisecond timescale, giving rise to correlated phase errors that may significantly affect the performance of quantum error correction protocols~\cite{kurilovich2025}.

Further evidence for the role of radioactivity has come from experiments performed in low-radioactivity environments. 
Operation in low-radioactivity facilities has been associated with reduced correlated error rates~\cite{Bratrud2024FNAL} and single qubit error rates~\cite{DeDominicis2026}, enhanced stability of fluxonium qubits~\cite{Gusenkova:2022}, and improved internal quality factors in superconducting resonators~\cite{Cardani:2021}. 

These findings have motivated several mitigation strategies aimed at improving the microscopic understanding of radiation-induced errors and suppressing their impact. 
On-chip approaches include phonon traps~\cite{Henriques:2019, martinis:2021, iaia_2022, Larson2025, lamagna2026phonondownconversionnormalmetals}, superconducting gap engineering~\cite{Marchegiani2022,kamenov2023gap_engineering, mcewen2024gap_engineering, Harrington:2024,pinckney2026,Binney2026}, and auxiliary particle detectors~\cite{orrell2021sensor,yang2026h}. 
Complementary efforts have focused on modeling the environmental background~\cite{cardani:2023,Fowler_2025,fowler2024spectroscopic,Loer_2024}, investigating the impact of different radiation environments by operating quantum processors at deep- or even shallow-underground sites~\cite{Bertoldo_2025} or at particle-beam facilities~\cite{mcjunkin2026,casagranda2025}, and identifying muon-induced events through integrated muon-tagging systems~\cite{Harrington:2024, Li2024direct, Mariani:2025, castelli2025superconductingqubitdecoherencecorrelated}.

Although radiation is typically regarded as a detrimental source of decoherence, the same physical processes suggest a complementary perspective: superconducting qubits are intrinsically sensitive to energy deposition events in their substrate. 
From this perspective, a quantum processor can be regarded as a spatially distributed, cryogenic calorimetric array.
Phenomenological studies of qubit sensitivity~\cite{Linehan_2025} suggest an energy threshold on the order of hundreds of eV for present-day transmons, and as low as 0.1\,eV for devices specifically engineered for this purpose. 
These projections have stimulated the development of novel superconducting sensors with unprecedented sensitivity~\cite{Fink:2024,magoon2026}, with potential applications in low-energy particle physics such as dark matter searches~\cite{essig2023snowmass2021cosmicfrontierlandscape} and coherent elastic neutrino–nucleus scattering~\cite{Cadeddu_2023}.
Quantitatively characterizing this response requires an independent measurement of the particle interactions in the qubit substrate. The first experimental study operating a qubit as a particle detector reported a detection efficiency below 10$\%$  for $\gamma$-rays depositing on average approximately 100 keV in the chip~\cite{DeDominicis2026}. In that experiment, however, the particle interaction rate was inferred from Monte Carlo simulations. As a consequence, the absolute detection efficiency could not be determined against an independent experimental reference.

Here, we investigate a superconducting multi-qubit processor as a cryogenic particle detector. 
By instrumenting the chip with an independent cryogenic calorimeter and exposing it to controlled radioactive and thermal excitations, we reconstruct particle-induced events directly from qubit signals and quantitatively characterize their detection capabilities. The calorimetric measurement provides, for the first time, an absolute determination of the qubit detection efficiency together with an experimentally constrained estimate of the effective energy threshold. 
Furthermore, by comparing the qubit response to particle interactions with that produced by controlled injections of thermal phonons, we show that the observed signals cannot be explained by transient substrate heating alone, supporting their origin in the non-equilibrium phonons generated by radioactive interactions. 
These results provide a quantitative characterization of superconducting qubits as cryogenic particle detectors and establish an experimental framework for connecting particle interactions with the underlying energy-deposition processes in superconducting quantum circuits.


\section{Experimental Setup}
\label{sec:setup}
The chip was fabricated on a 432\,µm-thick unannealed HEMEX-grade sapphire substrate measuring 7.5$\times$7.5\,mm$^2$, fully coated with a niobium film. Eight fixed-frequency niobium transmons with varying geometries, each incorporating standard Al/AlOx/Al Josephson junctions, were patterned on the substrate. 
To reduce losses associated with the formation of Nb$_2$O$_5$, the surfaces of the qubits were capped with a $\sim$10\,nm tantalum layer~\cite{Grassellino2023,alkhazaleh2026}.

The chip was instrumented with an independent thermal sensor consisting of a $3 \times 0.6 \times 0.4$\,mm$^3$ neutron-transmutation-doped germanium thermistor (NTD)~\cite{Haller}. 
The NTD was glued to the top side of the substrate. The energy released in the substrate raises its temperature, which is measured as a decrease in the NTD resistance.
NTD-based calorimeters achieve typical RMS energy sensitivity ranging from 20\,$\mathrm{eV}$ for 1\,$\mathrm{g}$ substrates~\cite{refLowMass} to 1.6\,$\mathrm{keV}$ for $2\,\mathrm{kg}$ substrates~\cite{refHighMass}. 
This sensor technology was selected because, in addition to its excellent sensitivity, it provides a calibrated measurement over the full keV--MeV energy range and operates over a broad temperature range (10--30\,mK), while imposing minimal constraints on the qubit chip design. Because of these characteristics, NTDs have been widely employed in particle physics~\cite{2025CUORE, Azzolini_2022, Augier_2022, 2025CUPID, Khalife_2020, edelweiss}.

In addition, a silicon heater was attached to the substrate to inject controlled thermal phonons~\cite{Alessandrello:1998bf, Andreotti:2012zz}.

Both the NTD and the heater were glued to the chip using the two-component epoxy Araldite\textsuperscript{\textregistered} Rapid, a standard choice for cryogenic detectors due to its excellent low-temperature performance and high radiopurity. A photograph of the assembled device is shown in Fig.~\ref{fig:setup}. 
Additional details on the NTD wiring, electronics, and readout chain are provided in Supplementary Section \ref{sec:ntd}.

\begin{figure}[t]
\includegraphics[width=\columnwidth]{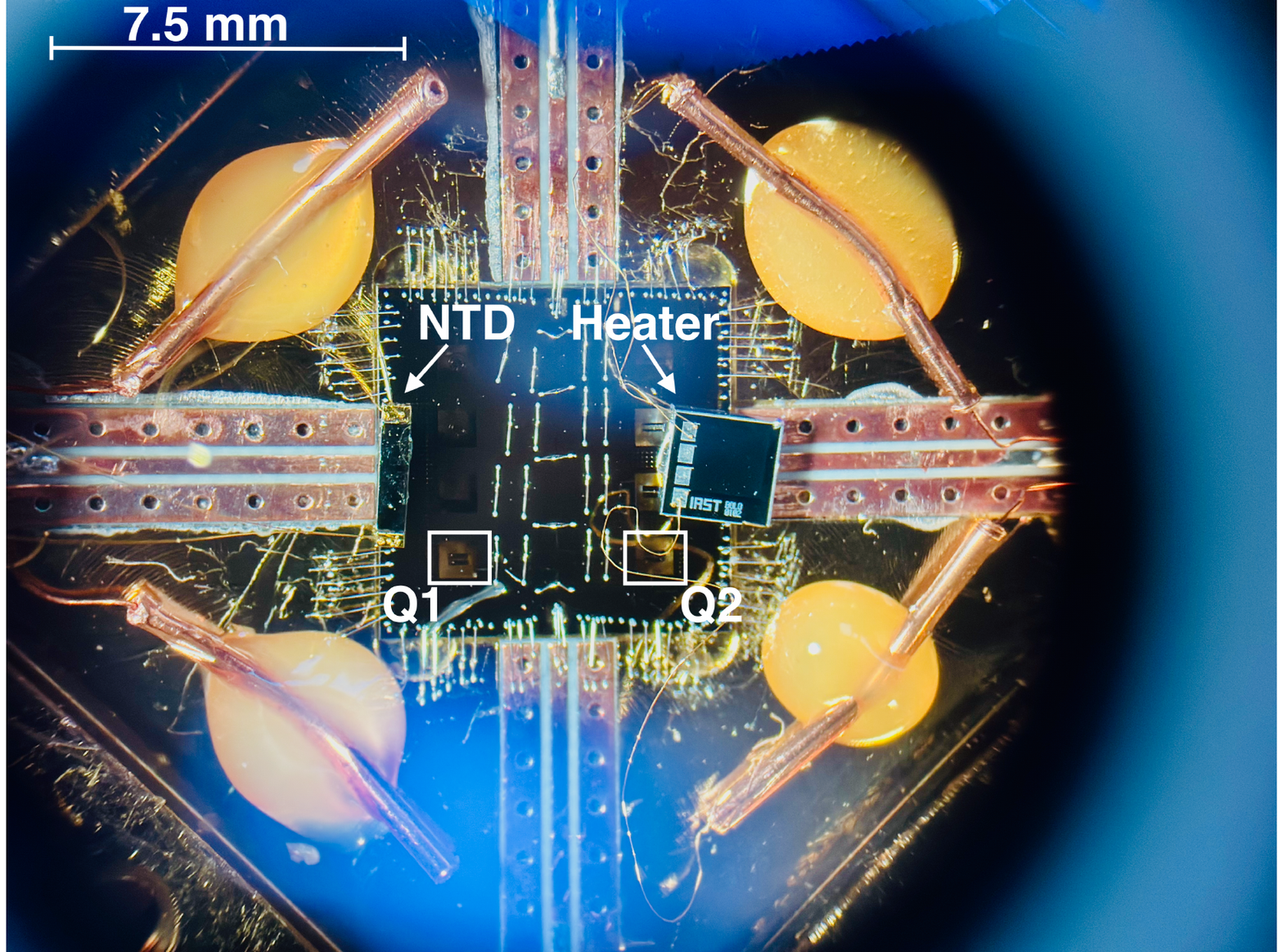}
\caption{Photograph of the assembled device, showing the qubit chip mounted in its copper holder and wire-bonded to the printed circuit boards. The chip is instrumented with an NTD thermistor and a Si heater. Arrows indicate the positions of the NTD and heater. }
\label{fig:setup} 
\end{figure}

The chip was continuously exposed to a radioactive source produced by implanting $^{224}$Ra nuclei onto a strip of adhesive tape attached inside the copper holder (Supplementary Section \ref{sec:source}).  
This isotope was selected for its relatively short decay constant ($\tau$=5.24\,days), which enables time dependent studies of the event rate.
The implanted $^{224}$Ra nuclei initiate a radioactive decay chain, whose daughter isotopes emit $\alpha$-particles, electrons, and $\gamma$-rays.
The energy spectrum resulting from source interactions in the chip was simulated using a Monte Carlo technique and is reported in Supplementary Section \ref{sec:source}.

The device was measured in the \href{https://ieti.sites.lngs.infn.it/}{``Ieti'' facility}, a dry dilution cryostat operated as an open-access research infrastructure and located in the underground INFN National Gran Sasso Laboratories (INFN--LNGS, L'Aquila, Italy). 
The 1.4\,km rock overburden reduces the muon flux by approximately six orders of magnitude.
The radiation shielding installed in the Ieti facility (comprising cryogenic and room-temperature lead and copper shields) suppresses the flux of $\gamma$-rays due to the natural radioactivity of the laboratory environment.
The resulting residual interaction rate in the chip was (5.3$\pm$0.5)$\times$10$^{-3}$\,events/s, one order of magnitude smaller compared to standard facilities.

Both qubit control and readout pulses were delivered through a common input line, attenuated and filtered at multiple temperature stages. The output signals were amplified by a travelling wave parametric amplifier (TWPA) at base temperature, followed by a commercial high-electron-mobility transistor (HEMT) amplifier at the 4\,K stage and a room temperature amplifier outside the refrigerator prior to digitization. 
Additional details on the qubit measurement setup are provided in Supplementary Section \ref{supp:readout}.

\section{Detection Protocol}
\label{sec:readout}
The qubits were operated using a ZCU216 RFSoC board controlled via the Quantum Instrumentation Control Kit (QICK)~\cite{stefanazzi2022qick}. 

We adapted the detection protocol introduced in Ref.\cite{DeDominicis2026} to simultaneously operate multiple qubits using multiplexed drive and readout pulses. The protocol relies on preparing each qubit in its first excited state $\ket{e}$ and measuring its state after a few $\mu$s, a time much shorter than the qubit's energy-relaxation time T$_1$. Under normal operating conditions, the qubit therefore has a high probability of being measured in $\ket{e}$. Particle interactions cause a transient suppression of T$_1$, increasing the decay probability and producing sequences of consecutive measurements in $\ket{g}$, which constitute the characteristic signature of the event. 

Qubit preparation at the beginning of each cycle is achieved by means of a conditional $\pi$-pulse that is applied when the qubit is found in $\ket{g}$, restoring it to $\ket{e}$. A cool-down period of a few tens of $\mu$s is included in each cycle to limit pulse-induced heating and unwanted thermal excitation.

To implement this protocol simultaneously on four qubits, we modified the QICK firmware by incorporating pre-existing blocks for the generation and readout of multiplexed signals~\cite{Ding_2024}. In addition, a custom configuration was developed featuring two multiplexed signal generators (one dedicated to drive pulses and the other to readout pulses) and a multiplexed readout module. 

The memory availability of the board limited the demodulation to four readout signals. Therefore, the present study is restricted to four qubits.
A key aspect in implementing the multiplexed qubit reset was the conditional application of the $\pi$-pulse. Since the four qubits may occupy different initial states, an unconditional $\pi$-pulse would drive unwanted transitions. To address this, the $\pi$-pulse was modified to include only the frequency tones corresponding to qubits found in $\ket{g}$. Additional implementation details of the multiplexed reset within QICK are provided in Ref.~\cite{FDD2026}.

\section{Particle-event identification}
\label{sec:analysis}
Data were collected at the underground INFN-LNGS laboratory in July 2025 using the multiplexed detection protocol described above. 
The analysis presented in the following is based on qubits Q1 and Q2, which provide stable operation and reliable state discrimination throughout the measurements. Details on the qubit selection are provided in Supplemental Section \ref{sec:details-selection}.

In each run, we acquired I/Q (in-phase and quadrature) signals for the qubits by executing the detection protocol with a sampling period of 60.7~$\mu$s. This value, primarily set by the cool-down time, corresponded to the fastest rate at which we could repeat the cycle without an excessive excitation of the qubit on the higher states.
Data traces (6 s duration) were fitted in the I/Q plane to determine state populations, excluding traces that fell outside target population requirements for the ground or first excited states (see Supplementary Section \ref{sec:details-selection}).
Although these selection criteria are stringent and significantly reduce the live time, they ensure a reproducible dataset and enable a robust comparison across runs acquired on different days.

For each trace and each qubit, a threshold in the I distribution was defined to discriminate between $\ket{g}$ and $\ket{e}$, thereby generating a binary stream of 0s and 1s with an average fidelity of $\sim 97-99\%$.

\begin{figure}[t]
\includegraphics[width=\columnwidth]{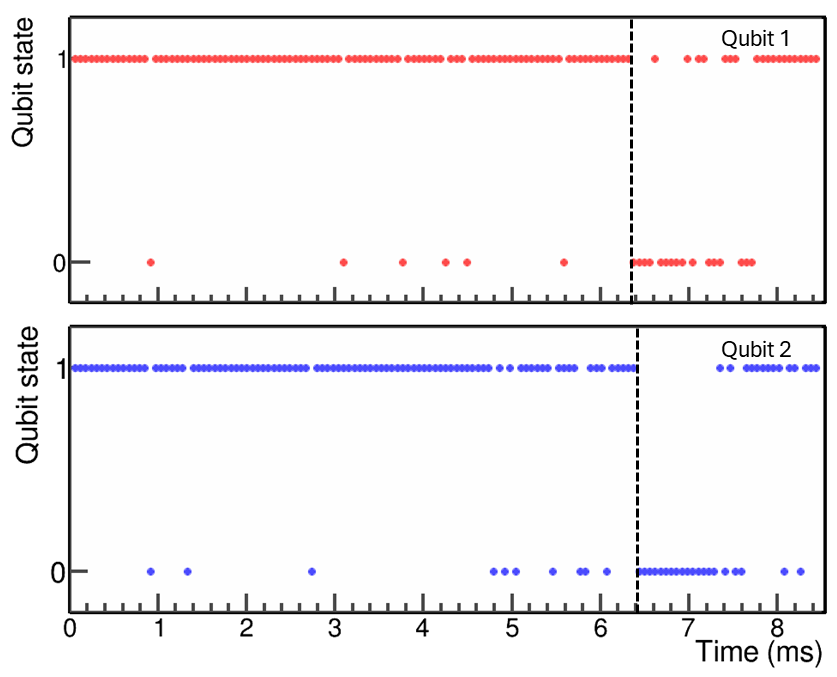}
\caption{Event simultaneously detected by two qubits. A window of 100 points preceding the trigger was also acquired to assess the state of each qubit prior to the interaction. The dashed black lines indicate the triggers position.}
\label{fig:event}
\end{figure}

Following the technique outlined in Ref.~\cite{DeDominicis2026}, a software trigger was run on the binary stream of each qubit, requiring four or more consecutive 0s to fire the trigger for that qubit. When the trigger fired, a window of $N_{\mathrm{signal}}=40$ points, corresponding to $\approx$2.4 ms, was recorded (the “signal region”). The size of this window was chosen such that it comprises the longest-lasting radiation-induced events observed. Additionally, $N_{\mathrm{control}}=$ \baseline\ preceding the trigger were stored (the “control region”) to assess the qubit state immediately before the event. This 140-point window (control + signal regions) is hereafter referred to as a candidate event. An example of an event in which two qubits detected a simultaneous radiation event is shown in Fig.~\ref{fig:event}.

Candidate events were further selected to reject spurious triggers, due to measurements instabilities, accounting for possible fluctuations in the qubits' T$_1$ ~\cite{Berritta:2025}.  
We estimated the likelihood that the number of 0s observed in the signal region cannot be explained by a fluctuation of the noise measured in the control region, i.e. \emph{signal probability} ($P_S$), whose explicit definition is reported in Supplementary Section \ref{sec:Psignal}. We computed $P_S$ for each candidate event and used it to perform a data selection. 
The threshold on $P_S$ was chosen to maximize the ratio between the signal selection efficiency, i.e. the fraction of signal events that pass the selection, and the background selection efficiency. This technique, commonly used in particle physics ~\cite{azzolini2018, lowe}, is described in details in Supplementary Section \ref{sec:Pscan}. The achieved signal survival efficiency is namely \effone\ and \efftwo\ for qubits Q1 and Q2, respectively.

As outlined in Supplementary Section \ref{sec:Pscan}, this selection technique was tested on data published in Refs.~\cite{DeDominicis2026}, yielding an efficiency improvement larger than $40\%$ compared to that work.
In the following sections we discuss the results obtained using the data selected through this procedure.

\section{Results}
\label{sec:results}
The independent measurement of the source interaction rate provided by the NTD allowed us to determine the absolute detection efficiency of the qubits.
To this end, we used runs collected between July 23 and July 30 2025, during which the activity of the $^{224}$Ra source decreased significantly because of its short decay constant ($\tau$=5.24\,days).

\begin{figure}[t]
\includegraphics[width=\columnwidth]{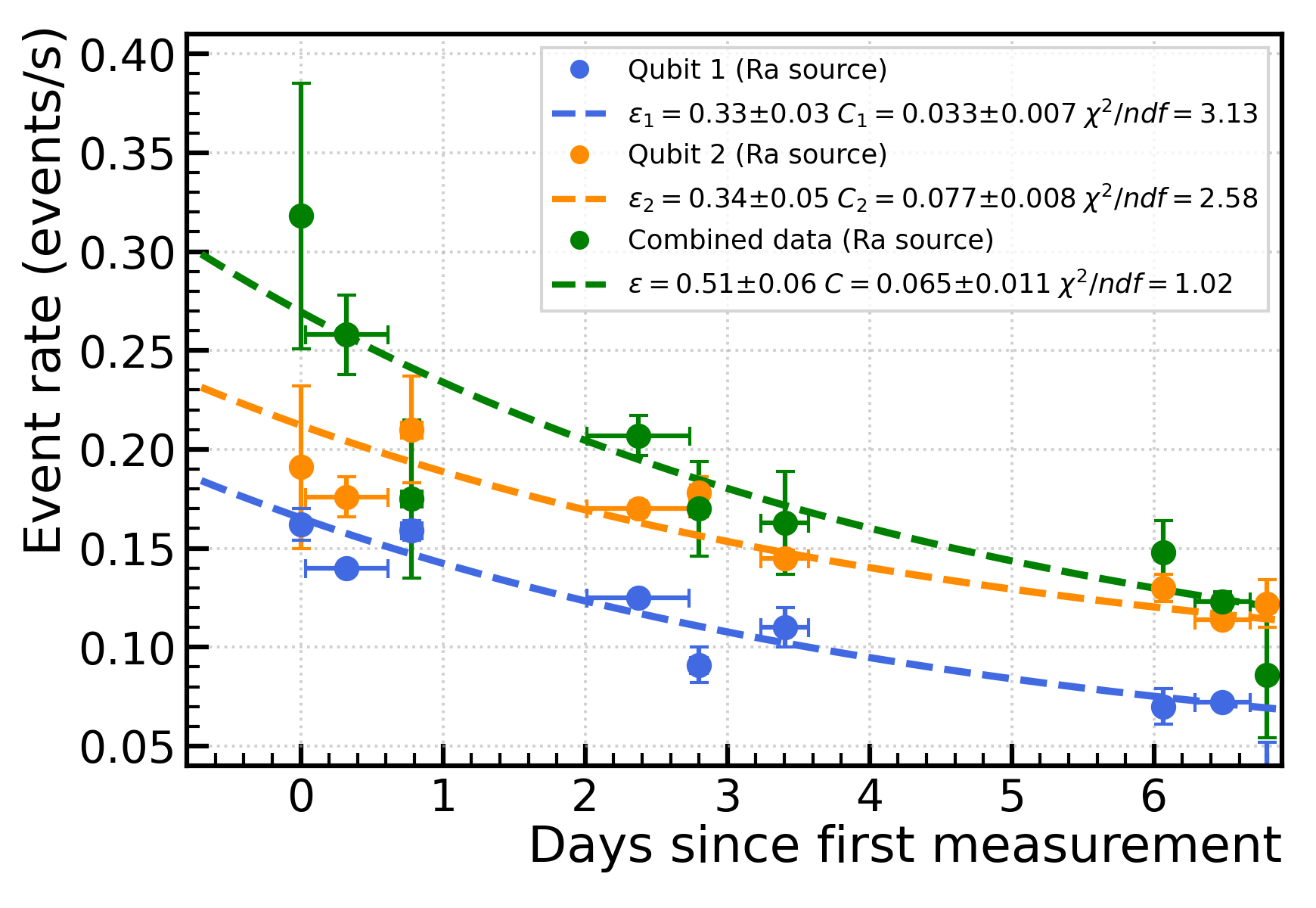}
\caption{Rate of selected events as a function of time since the start of the measurements. The blue, orange and green circular markers represent the data from Q1, Q2, and their combination, respectively. Horizontal uncertainties indicate the measurement duration; vertical uncertainties are derived from Poisson statistics propagated to the rate.
The dashed curves show exponential decay fits to the data, modeled using the absolute activity ($A_0$ =0.40 $\pm$ 0.01\,events/s), the decay constant of the $^{224}$Ra source ($\tau$=5.24\,days), the qubit detection efficiencies ($\varepsilon$), and a residual event rate (C), namely $A_0\varepsilon_ie^{-t/\tau}+C_i$. }
\label{fig:decay}
\end{figure}

Fig.~\ref{fig:decay} shows the rate of selected events as a function of time for qubits Q1 and Q2. 
The data were modeled using an exponential decay function with the decay constant fixed to that of the radioactive source, plus a constant term accounting for environmental radioactivity and residual noise $C$. 
The exponential function was multiplied by a constant factor, $A_0 \times \varepsilon_i$, where $A_0$ is the initial impact rate from the source in the substrate and $\varepsilon_i$ is the detection efficiency of the $i$-th qubit. 
The simultaneous operation of the NTD with the qubits allowed us to reconstruct the rate of particles hitting the substrate at any time of the measurement. $A_0$ was thus estimated using the NTD, giving $A_0= (0.40\pm0.01)$\,events/s. (Supplementary Section \ref{sec:source}.)

The reference fit reported in Fig.~\ref{fig:decay} showed that the model provides a good description of the data, with inferred detection efficiencies of $\sim$33$\%$ and $\sim$34$\%$ on Q1 and Q2, respectively. 
The fit was repeated under different conditions to assess the systematic uncertainty associated with alternative trigger requirements, different signal-region lengths, and different initial selections of traces.
The resulting efficiency values were averaged, and their spread was taken as an estimate of the systematic uncertainty. 
The corresponding efficiencies were $(31_{-6}^{+5})\%$ and $(39_{-11}^{+8})\%$ for qubits Q1 and Q2, respectively. 

We then investigated whether the spatially distributed response of the device could be exploited to improve the detection efficiency.
Indeed, events that produce a weak signal in a qubit could be partially covered by noise (Fig.~\ref{fig:histo_zero}). Exploiting a second qubit triggered in coincidence would allow us to amplify the signal while suppressing contributions from random noise. We investigated the potential of this approach by using the most sensitive qubit, Q2, as a trigger. When Q2 triggered, we searched for a coincident signal in Q1 and summed the 0s outputs of the two qubits. The data selection described in the previous paragraph was then repeated using the \emph{combined} signal.
Combining Q1 and Q2 increased the detection efficiency to $(51_{-10}^{+11})\%$ (green dots in Fig.~\ref{fig:decay}), demonstrating that a multi-qubit architecture can substantially increase the probability of detecting an interaction in the substrate. Details of the coincidence analysis are provided in the Supplementary Section~\ref{sec:combination}.

Despite this improvement, approximately half of the particle interactions remained undetected even by the combined two-qubit system.
These missing events are unlikely to be lost because of random noise fluctuations, since the selected event population is efficiently separated from the background (see Fig. \ref{fig:scanP}).
Instead, the incomplete detection efficiency suggests that some particle interactions do not produce a measurable qubit response. This behavior may arise from several physical mechanisms, including a minimum deposited energy required to perturb the qubit or the spatial dependence of the interaction with respect to the qubit location. Recent work has shown that these effects are closely interconnected and cannot be easily disentangled on an event-by-event basis~\cite{celi2026}. 
Nevertheless, the measured inefficiency can be used to place an upper limit on an energy scale below which particle interactions may remain undetected. We considered the conservative limiting case in which the entire loss of efficiency was attributed to an energy threshold, neglecting any spatial dependence of the qubit response.

\begin{figure}[t]
\includegraphics[width=\columnwidth]{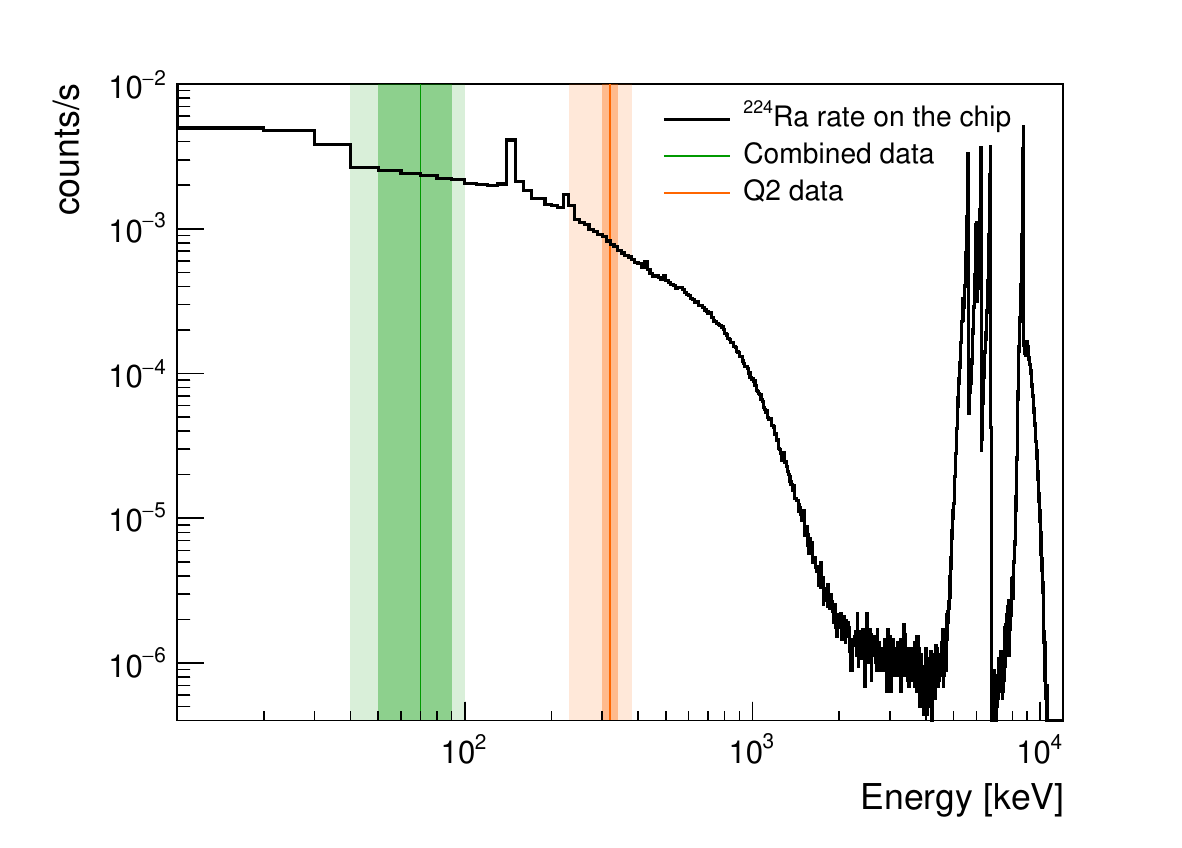} 
\includegraphics[width=\columnwidth] {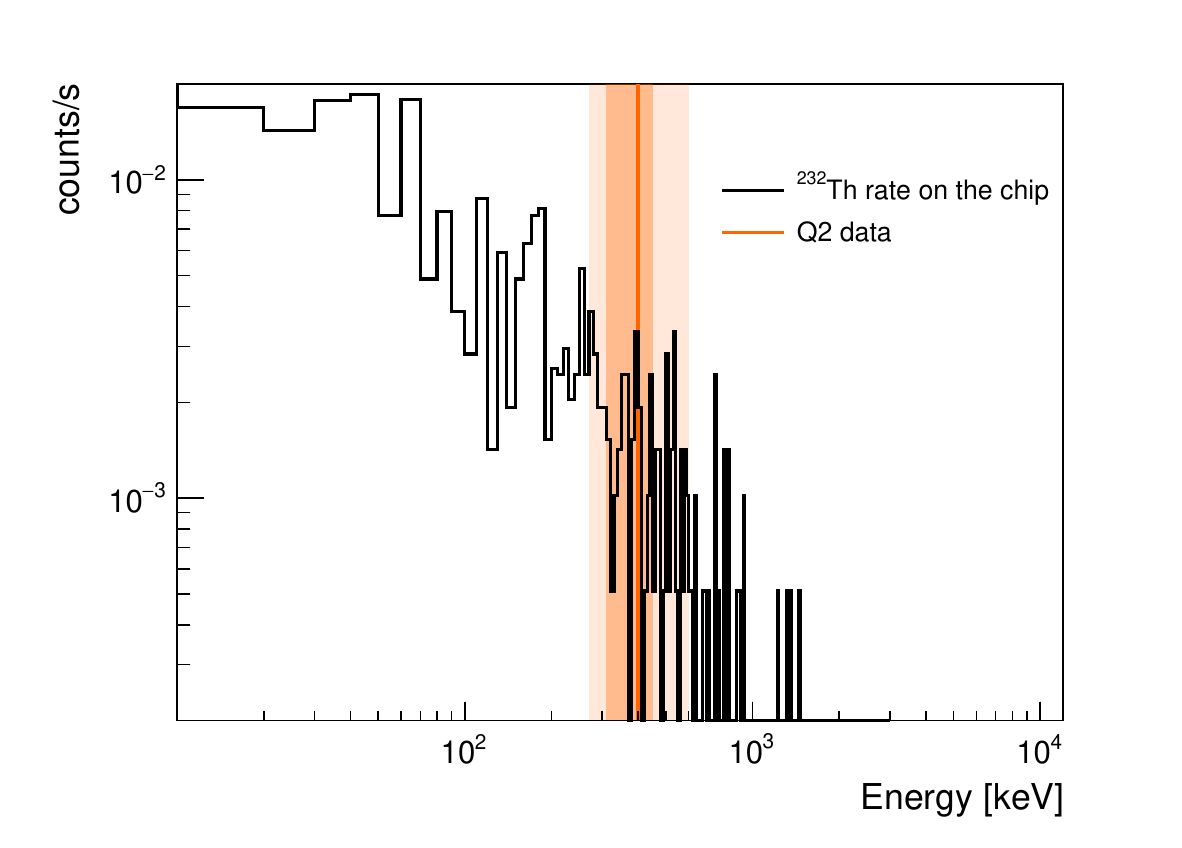}
\caption{Top panel: Monte Carlo simulation of the $^{224}$Ra spectrum weighted by the particle hits rate as reconstructed by the NTD. In orange and green the energy threshold for a single qubit and the combination of two, respectively, estimated from the corresponding efficiency. Bottom panel: simulated $^{232}$Th external source spectrum with the estimated single qubit threshold. The dark and light shaded areas represent the statistical and systematic uncertainty on the threshold. }
\label{fig:threshold} 
\end{figure}

For this purpose, we normalized the Monte Carlo spectrum of the energy deposited by the source in the chip to the total number of interactions independently measured with the NTD sensor.
Then, we integrated this spectrum from high to low  energy. The energy at which the cumulative rate equals the rate measured by the qubit is taken as the effective energy threshold (Fig.~\ref{fig:threshold}).

We performed this study on a run with high statistics. 
This procedure gives a threshold of $E_{threshold}^{Q1+Q2}=70\pm20~\mathrm{(stat)}\pm30~\mathrm{(syst)}~\mathrm{keV}$. 
For comparison, applying the same procedure to Q2 alone gives $E_{threshold}^{Q2}=320\pm20~\mathrm{(stat)}^{+60}_{-90}~\mathrm{(syst)}~\mathrm{keV}$.
We stress that this threshold should be regarded as an effective threshold, since it relies on the assumption that the detection efficiency is entirely determined by the deposited energy. 
These values should therefore be interpreted as upper limits on an effective energy threshold, rather than as intrinsic sharp thresholds of the qubits.
While the achieved threshold is orders of magnitude above the state-of-the-art of low threshold cryogenic detectors~\cite{edelweiss,cresst,tesseract} operating different sensors, it must be underlined that the present device was designed as a quantum processor rather than as a particle detector. The measured threshold therefore provides a baseline for conventional transmon devices rather than a limit on the sensitivity achievable with qubits specifically engineered for particle detection.

In order to have an independent measurement of the effective threshold, we deployed a second radioactive source with substantially different energy spectrum. To this aim, we placed a $^{232}$Th source in the proximity of the cryostat. 
The $^{224}$Ra source installed inside the chip produced approximately $60\%$ $\alpha$-particles with energies from 5 to $9~\mathrm{MeV}$, while the remaining emissions arise from electrons and $\gamma$-rays, mostly below $2.6~\mathrm{MeV}$. 
On the contrary, the $\alpha$-particles and electrons emitted by the $^{232}$Th source are fully absorbed by the cryogenic apparatus, while part of the $\gamma$-rays reach the chip, depositing energy in the substrate. Thus, the majority of events produce only a few tens to a few hundreds of keV.
These considerations, supported by Monte Carlo simulations and by the energy spectrum reconstructed with the NTD, indicate that, due to the presence of $\alpha$ particles, energy deposits from the internal $^{224}$Ra source are, on average, larger than those produced by the external $^{232}$Th source (Fig. \ref{fig:sim_spectra}). 

For this analysis we used only Q2 data, as Q1 was affected by grounding issues in the NTD readout electronics during this measurement. 
We obtained an efficiency of $\sim 23\%$, lower than the efficiency measured with the more energetic $^{224}$Ra source ($\sim 39\%$), as expected.
The reduced efficiency observed for the softer spectrum provides direct experimental evidence that the qubit response depends also on the energy deposited in the substrate.

Assuming once again that the detection efficiency is determined solely by the deposited energy, this efficiency corresponds to an energy threshold of:
$E_{threshold - \gamma}^{Q2} = 400^{+200}_{-130}~\mathrm{(stat)}^{+50}_{-90}~\mathrm{(syst)}~\mathrm{keV}$ (Fig.~\ref{fig:threshold}-bottom). 
This provides an independent cross-check of the effective threshold estimate.
Despite the different sources and detection efficiencies, the inferred thresholds are consistent within uncertainties.

This threshold estimation method was based on the assumption that the qubit response was determined only by the deposited energy.
While this assumption provided a convenient figure of merit for quantifying the sensitivity of the qubits to particle interactions, other phenomena have to be considered in order to give a full picture of the underlying physics. 
This is supported by a complementary measurement in which we investigated whether purely thermal energy deposition could reproduce the response observed for particle interactions.
For this purpose, thermal phonon pulses were injected into the qubits substrate using the silicon heater (Fig.~\ref{fig:setup}).
In detail, we injected 15-millisecond-long thermal pulses at 10 Hz over the 1–11 MeV energy range, with their deposited energy independently reconstructed by the NTD.
In detail, we injected  thermal pulses with energies from 1 to 11\,MeV and with frequency of 10\,Hz.
Based on the substrate heat capacity, the maximum deposited energy corresponded to an equivalent transient temperature increase of approximately 240\,mK.

As a first test, we searched for a thermal response of the qubit by monitoring its excited-state population as a function of the injected heater energy. 
Superconducting qubits have recently been studied as thermometers under near-equilibrium conditions~\cite{Lvov:2025}. 
Motivated by these results, we investigated whether the heater pulses produced a measurable increase in the qubit excited-state population. However, no statistically significant variation was observed over the entire explored energy range.

\begin{figure}[t]
\includegraphics[width=\columnwidth]{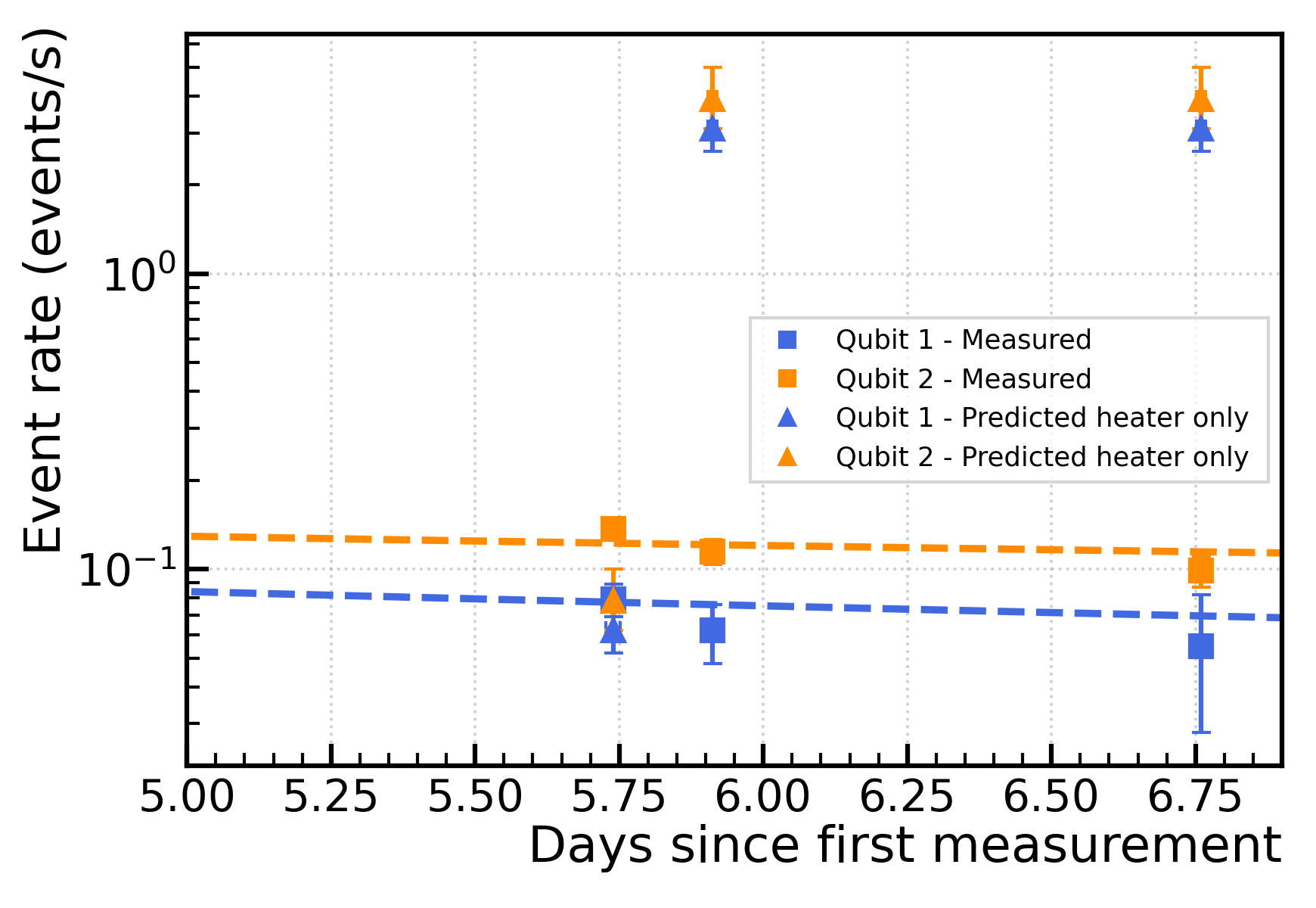}
\caption{Blue and orange curves: fits to the event rates induced by the radioactive source as a function of time (Fig.~\ref{fig:decay}). Triangles: expected event rates produced by the heater in Q1 (blue) and Q2 (orange), as determined from the NTD measurement and corrected for the corresponding single-qubit detection efficiencies. Squares: event rates measured by Q1 (blue) and Q2 (orange). The measured rates are consistent with the contribution expected from the radioactive source alone, with no significant excess attributable to the heater. }
\label{fig:heater} 
\end{figure}

Since the absence of a change in the average qubit population did not exclude the possibility of transient responses associated with individual heater pulses, we further analyzed the data using the same event-selection algorithm developed for particle-induced events. Again, no statistically significant excess of events was observed in coincidence with the heater pulses (Fig.~\ref{fig:heater}).

These two complementary measurements showed that a transient thermal excitation of the substrate was not sufficient to reproduce the observed qubit response, even when the deposited energy was comparable to or larger than that released by particle interactions.
Instead, the measurements were consistent with the interpretation that the signals reported in this work originated from the non-equilibrium phonon population generated by particle interactions, rather than from the transient increase in substrate temperature alone.

\section{Conclusions and Perspectives}
In this work, we investigated a superconducting multi-qubit chip as a cryogenic particle detector by combining multiplexed qubit readout with an independently instrumented calorimetric sensor. 
This approach provides an experimental reference for the absolute particle interaction rate, allowing us to directly measure qubit detection efficiencies and constrain their effective energy sensitivity.

The simultaneous operation of qubits and a calibrated heater also showed that, while particle interactions were efficiently detected by the qubits, heater pulses depositing comparable or larger amounts of energy do not produce a measurable response. 
These observations supported the interpretation that radiation-induced events were due to the non-equilibrium phonon population, rather than to the transient increase in the substrate temperature alone.

The approach presented here is complementary to recent methods that reconstruct particle interactions exclusively from the dynamics of superconducting qubits~\cite{celi2026}. 
While those approaches provide detailed information on the spatial and temporal evolution of particle-induced quasiparticles, the independent calorimetric measurement introduced in this work provides an absolute reference for the interaction rate and deposited-energy spectrum. 
Combining these complementary approaches offers a promising route toward a complete characterization of radiation-induced processes in superconducting quantum circuits as well as the exploration of superconducting qubits as cryogenic particle detectors for low-energy rare-event searches.

Finally, we note that the detection protocol adopted in this work was designed to maximize the efficiency of radiation-event identification rather than to reconstruct the energy of individual interactions. 
Indeed, particle interactions spanning a broad range of deposited energies (keV -- MeV) appear to produce signals of comparable duration in the qubit response (with an average exponential decay of 0.7 millisecond). As a consequence, a particle event generally results in a stream of 15--20 zeros, which does not show a clear correlation with the deposited energy. 
While this is sufficient for the quantitative characterization performed here, it also highlights a limitation of the current approach for applications requiring event-by-event energy reconstruction or particle spectroscopy. In this context, recent approaches that exploit the full temporal evolution of the qubit response to reconstruct quasiparticle dynamics represent a promising and complementary direction toward improving the energy resolution of superconducting-qubit-based particle detectors~\cite{Larson2025,sundelin2026}.

Together, these developments establish a route toward using superconducting qubit arrays both to quantitatively characterize radiation-induced processes in quantum devices and as distributed cryogenic particle sensors.
\section*{Acknowledgements}
The authors thank the Director and technical staff of the Laboratori Nazionali del Gran Sasso. We are also grateful to the LNGS Computing and Network Service for computing resources and support on U-LITE cluster at LNGS. We thank C.~Tomei and G.~D'Imperio for the maintenance of the simulation software and we are grateful to A.~Girardi for the controller of the cryogenic switch, M.~Guetti for the help with the cryogenic facility (including its continuous upgrades), and M.~Iannone for the assembly of the device at LNGS. 
This work was supported by the Italian Ministry of Foreign Affairs and International Cooperation, grant number US23GR09, by the PNRR MUR project CN00000013-ICSC, and
by the Italian Ministry of Research under the PRIN Contract No. 2022BP4H73 and under the PRIN Contract No. 202273p73a.
This work was supported in part by the U.S. Department of Energy, Office of Science, National Quantum Information Science Research Centers, Superconducting Quantum Materials and Systems Center (SQMS), under Contract No. 89243024CSC000002 (MB, FC, SG, TY, SZ and AG).\\

\onecolumngrid

\section*{References}

\twocolumngrid
\bibliography{bib_v4}

\appendix
\section*{SUPPLEMENT}
\beginsupplement
\begingroup
\renewcommand{\thesubsection}{\Alph{subsection}}

\subsection{Data acquisition and analysis of the NTD}
\label{sec:ntd}
The qubit substrate was equipped with a germanium neutron-transmutation-doped thermistor (NTD)~\cite{Haller}, measuring 3 $\times$ 0.6 $\times$ 0.4 mm$^3$, glued to the upper surface with three spots of bi-component epoxy resin (Araldite\textsuperscript{\textregistered} Rapid). NTDs are thermal sensors obtained by irradiating high-purity germanium crystals with neutrons. The neutron-transmutation process introduces an extremely homogeneous dopant concentration, bringing the material close to the insulator-metal transition. In this regime, the electrical transport is described by the Mott Variable Range Hopping model~\cite{ShklovskiiEfros1984}, according to which charge carriers move through phonon-assisted hopping between localized states. As a consequence, the resistance of the thermistor exhibits a strong dependence on temperature. 
The NTD is biased with a constant current source. 
Energy deposited in the substrate causes a temperature rise, changing the NTD resistance and, under constant-current bias, the voltage drop across the sensor.
Therefore, the measured voltage pulse provides a direct estimate of the energy deposited in the substrate. The NTD are connected to the bias and readout system using 25~$\mu$m gold wires. These wires are bonded to twisted pairs of constantan wires, which are routed to an interface board mounted on the mixing-chamber plate of the cryostat. The signals are then transmitted to room-temperature electronics comprising low-noise front-end amplifiers, DC-coupling stages, and 24-bit ADC digitizers~\cite{Arnaboldi:2004,Arnaboldi_2018,Carniti:2022coa}.

The output passed through a 500~Hz anti-aliasing Bessel filter to suppress high-frequency noise. The resulting waveforms are synchronously recorded as a continuous data stream using a MATLAB-based data acquisition system~\cite{Carniti:2022coa} with a sampling frequency of 15.6~kHz.
The continuous data stream is processed offline using the C++ analysis framework \texttt{Octopus}~\cite{octopus}. 
Signal events are identified with a software trigger that activates when the waveform amplitude exceeds eight times the baseline RMS, where the baseline is calculated using a 4~ms window preceding the trigger. In addition, noise events are periodically recorded every 5~s to monitor the detector noise and provide reference waveforms for the signal processing.
Signal and noise templates are selected from the triggered events to construct an optimum filter~\cite{Gatti:1986cw}, which maximizes the signal-to-noise ratio of the recorded pulses. The filtered waveforms are then subjected to a series of pulse-shape quality cuts to reject non-physical events and poorly reconstructed signals.

This analysis chain was applied to all acquired datasets. In total, six datasets were collected to measure the $^{224}$Ra decay rate, eleven datasets were acquired with different heater pulse amplitudes to characterize the detector response. 
The calibration campaign with the $^{224}$Ra source lasted approximately six days, providing a sufficiently long observation period to reconstruct both the source activity and its half-life. For each dataset, the event rate was determined and plotted as a function of time, as shown in Fig.~\ref{fig:rate}. 

\begin{figure}[t]
\includegraphics[width=\columnwidth]{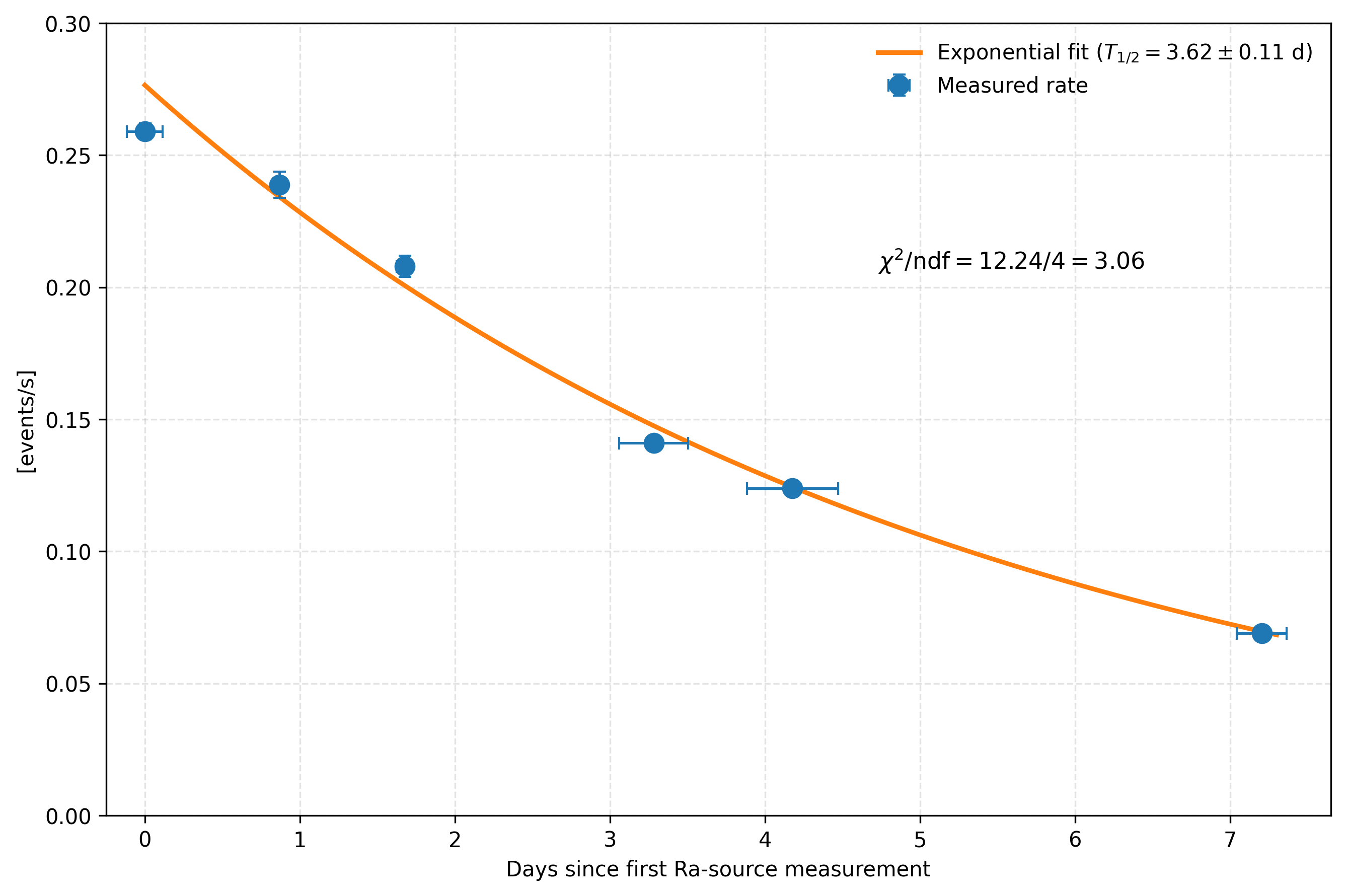}
\caption{Rate measured by the NTD as a function of time since the first day of data taking. The curve was fitted with an exponential decay (red curve), obtaining a decay constant compatible with the half-life of the $^{224}$Ra source.} 
\label{fig:rate}
\end{figure}

The detector energy scale was calibrated using the $\alpha$ peak produced by the $^{224}$Ra source. The trigger efficiency and analysis threshold were determined from dedicated heater scans performed by injecting pulses corresponding to different equivalent deposited energies ($\sim$20 keV steps from 0 to 100\,keV, then 100\,keV steps up to 1\,MeV, then 3.5 and 11\,MeV). 
The trigger efficiency analysis allowed us to set an analysis threshold of 96~keV where the NTD has an efficiency of 50\%. 
We emphasize that the performance achieved in this work is not representative of the intrinsic capabilities of NTD-based detectors. The relatively high analysis threshold is primarily a consequence of the unoptimized experimental setup and the elevated vibration-induced noise present during the measurements.

\subsection{Radioactive Sources}
\label{sec:source}
The radioactive source was produced by implanting $^{224}$Ra nuclei onto a strip of adhesive tape, which was positioned facing both the NTD and the qubits during the characterization measurements. 

The implantation was achieved by placing the tape at a distance of approximately 1\,cm from a primary $^{228}$Ra source inside a pumped vacuum chamber, following the same recoil‑based collection mechanism described for secondary‑source production in Ref. \cite{biassoni2022}. 
When $^{228}$Th undergoes alpha decay within the primary source, the resulting nuclear recoil imparts sufficient kinetic energy to the daughter nuclei to allow $^{224}$Ra to escape the substrate and implant onto the facing tape surface. The tape was exposed for a duration exceeding several half-lives of $^{224}$Ra, ensuring that equilibrium was reached between the implantation rate and the radioactive decay rate. This procedure yields an ultra‑shallow active layer, only a few tens of nanometers thick, suitable for experiments requiring minimal self‑absorption and direct geometric coupling between the alpha‑emitting nuclei and the qubit devices. 
An additional advantage of this method is that, after a few half‑lives, both the source and the qubit return to a negligible residual activity.

The energy deposited in the chip by the $^{224}$Ra source decay products has been simulated with a Geant4 based Monte Carlo algorithm. The relative energy spectrum is shown in Fig. \ref{fig:sim_spectra} in comparison to the one expected when placing a $^{232}$Th source outside of the cryostat. 

\begin{figure}[t]
\includegraphics[width=\columnwidth]{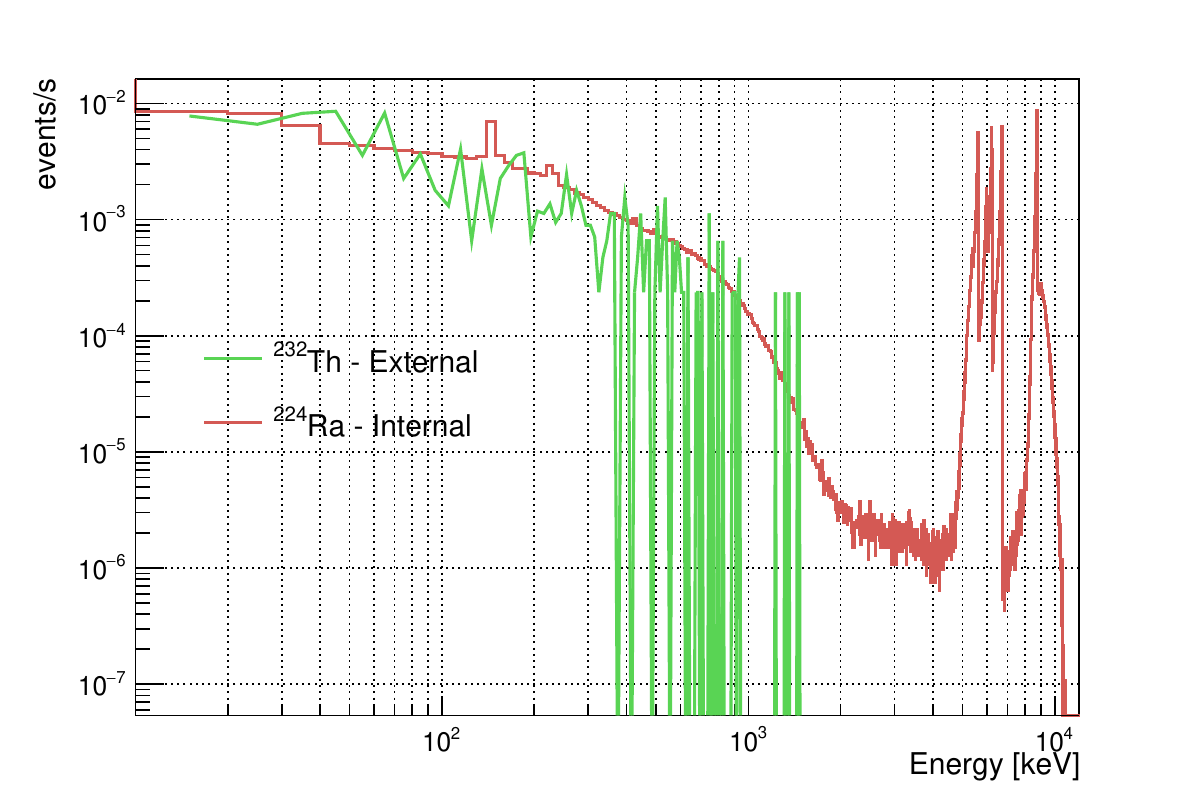}
\caption{Energy spectra, simulated with Geant4, expected by radiation impacting on the chip, due to a close $^{224}$Ra source at the beginning of the measurements (in red) and a 142\,kBq $^{232}$Th source placed outside of the cryostat (in green). }
\label{fig:sim_spectra}
\end{figure}

The initial rate of events impacting the qubit chip was measured with the NTD to be $\sim$0.47\,events/s. 
However, spurious events with energies close to the NTD threshold could be occasionally triggered, resulting in a higher rate. To correct for these effects, we only used the NTD spectrum at high energies, where pulses are not affected by noise and are reconstructed with full efficiency. We computed the measured rate using these events, and then extrapolated the value to the full energy range using the Monte Carlo simulated spectra (Fig.~\ref{fig:sim_spectra}).
For example, in the case of the $^{224}$Ra source we measured the NTD event rate above 3\,MeV and then extracted the total rate using the Monte Carlo simulation. The calculation was repeated starting from different values around $3~\mathrm{MeV}$ and provided consistent results. We also tested the goodness of this approach by estimating the rate at different energies well above the analysis threshold (400\,keV, 500\,keV) and ensuring that the obtained rate was consistent with the one measured by the NTD.
After this correction, the initial interaction rate is estimated as $A_0= (0.40\pm0.01)$\,events/s.

\subsection{Readout of the qubits}
\label{supp:readout}
Each of the qubits of the chip employed was capacitively coupled to a LC resonator connected to the same feedline. Qubit state readout was performed by means of dispersive shift measurements. To simultaneously probe multiple qubits, readout pulses with multiple carrier frequencies, at the resonant frequencies of the corresponding readout resonators, were employed. Pulses were generated by an AMD Zynq\textsuperscript{TM} Ultrascale+\textsuperscript{TM} RFSoC ZCU216 board equipped with the QICK firmware \cite{Ding_2024}. For this experiment, the QICK firmware was customized to feature two multiplexed generators, one for the drive and the other for the readout pulse, and one multiplexed readout \cite{FDD2026}. The multiplexed generators are capable of generating signals with up to eight tones. The readout, however, can demodulate only pulses with up to four tones, limiting the number of qubits that can be probed simultaneously. Pulses for the control and readout of the qubits reached the device after being attenuated at different temperature stages and filtered to stop infrared radiation from higher temperature stages. Transmitted readout signals were amplified in three different stages to maximize readout fidelity. The first amplification stage was already at 10 mK by means of a Carthago traveling wave parametric amplifier (TWPA) produced by Silent Waves \cite{Ranadive2022}. Following that, the signal is amplified further at 4K with a LNF-LNC4\textunderscore8G High Electron Mobility Transistor (HEMT) amplifier produced by Low Noise Factory. Finally, the last amplification stage was at room temperature with a Miteq AFS3-04000800-09-CR-4 amplifier. A complete scheme of the RF components employed for this measurement is portrayed in Fig. \ref{fig:rf_scheme}.

\begin{figure}
    \centering
    \includegraphics[width=\linewidth]{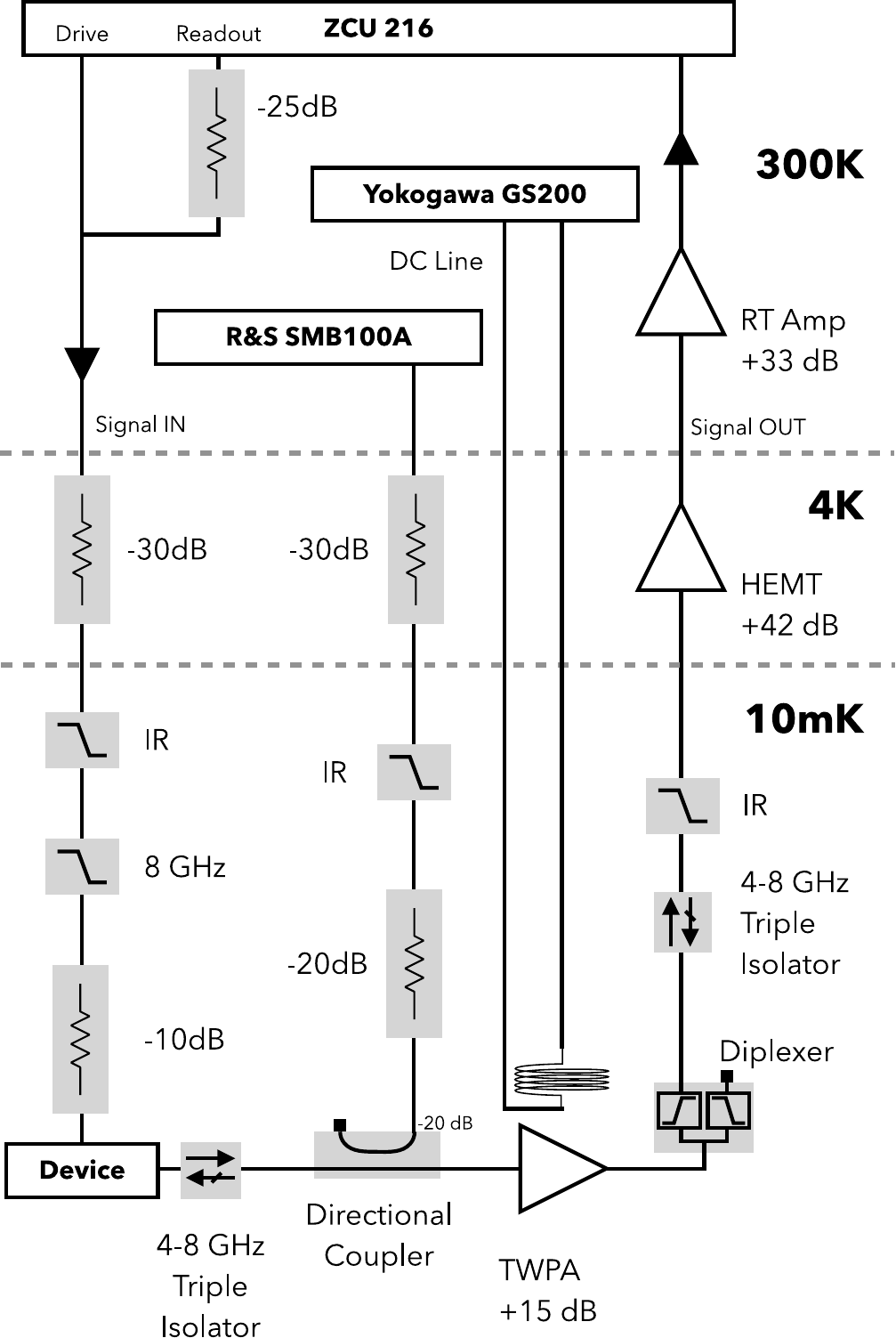}
    \caption{Scheme of the RF lines and components used for this experiment.}
    \label{fig:rf_scheme}
\end{figure}

\subsection{Traces Selection}
\label{sec:details-selection}
For each qubit, we fitted the clouds in the I/Q plane to determine the fraction of events in the ground state $\ket{g}$, as well as in the first $\ket{e}$ and second $\ket{f}$ excited states. The fit was performed on traces corresponding to $\sim$6\,s, or $10^5$ events, in order to monitor the level of populations as a function of time. In most of the data sets we found three clouds and we fitted them with  a sum of three two-dimensional Gaussian functions. The functions parameters are independent for each population.  

The description of the fit procedure is reported in the Supplemental Material of Ref. \cite{DeDominicis2026}. The result of this procedure for a typical run is shown in Fig.~\ref{fig:clouds_fit}-top.

\begin{figure}[t]
\includegraphics[width=\columnwidth]{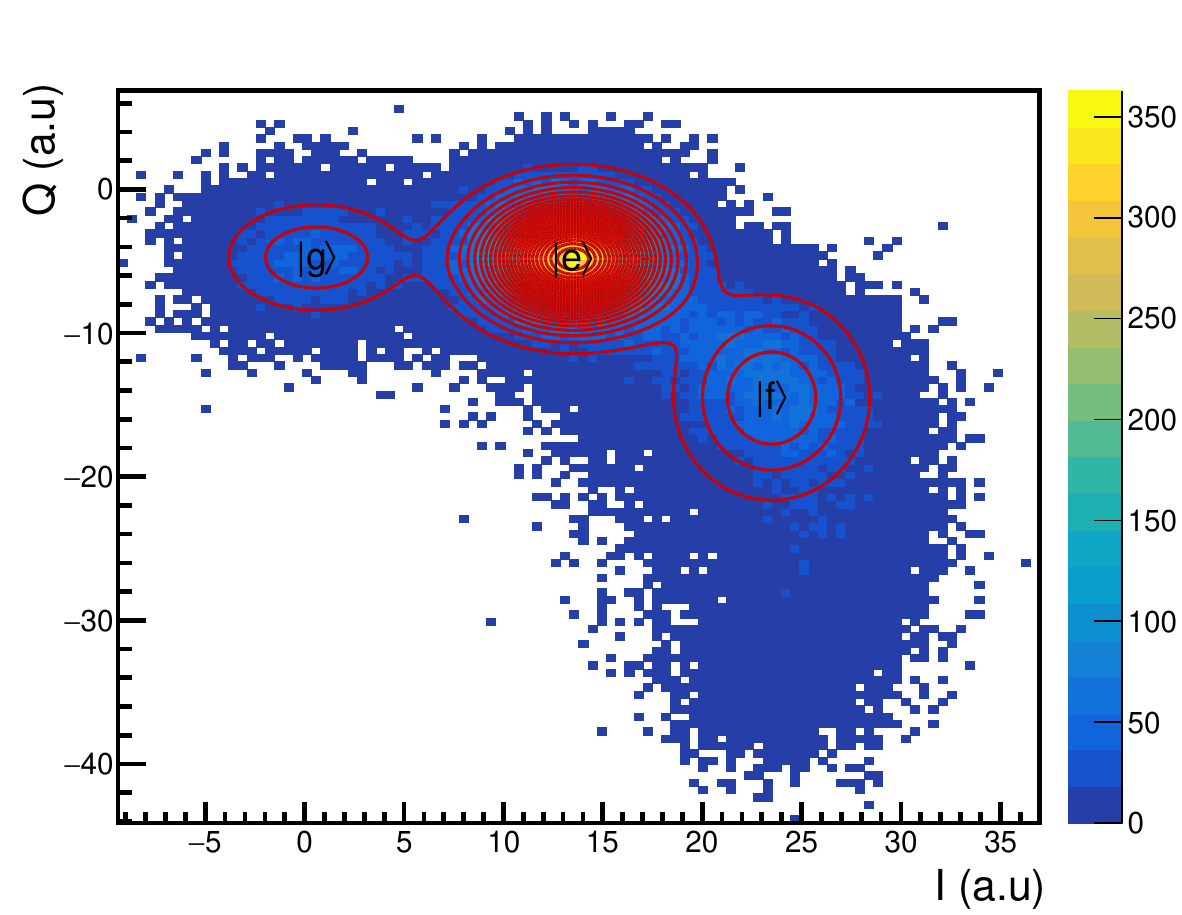}
\includegraphics[width=\columnwidth]{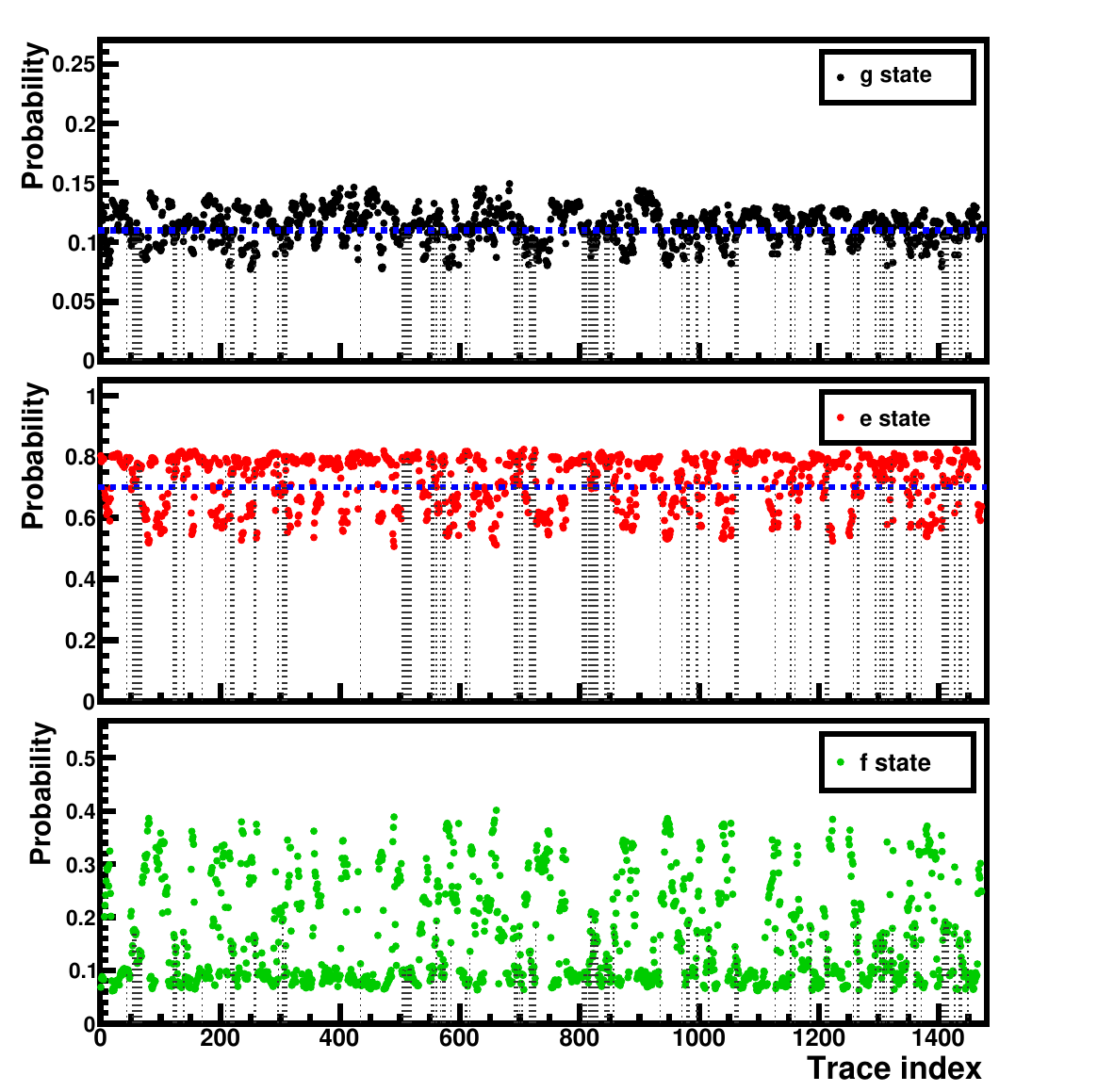}
\caption{Example of the state discrimination procedure for qubit Q1 in the I/Q plane. The upper plot shows three blobs corresponding to the qubit ground ($\ket{g}$), first excited ($\ket{e}$), and second excited ($\ket{f}$) states. These blobs are fitted with a sum of three two-dimensional Gaussian distributions (red contour lines). The fit is performed over $10^5$ samples. The population of each state is estimated from the fitted amplitudes and covariance parameters of the corresponding Gaussian components. The lower plot shows the extracted qubit states populations for each trace in a dataset containing 1470 traces. The blue dotted lines report the upper and lower cut for $\ket{g}$ and $\ket{f}$ probability, respectively. The traces selected with this criterion are indicated with grey dotted lines. }
\label{fig:clouds_fit}
\end{figure}

Q3 exhibits a large fraction of traces in which the $\ket{f}$ and $\ket{h}$, second and third excited states, respectively, are significantly populated, preventing reliable discrimination of radiation-induced relaxation events. Q4 showed significantly degraded performance compared with its previous characterization~\cite{Grassellino2023} and was therefore excluded.

For Q1 and Q2, we selected only the traces satisfying the population requirements $P(g) < 0.11$ and $P(e) > 0.7$.
A small value of $P(g)$ ensures that the number of zeros produced by noise is kept under control. A large value of $P(e)$ is required because, in our simple model, phonons cause the qubit to decay from $\ket{e}$ to $\ket{g}$. If higher excited states are populated, this model no longer applies. The selected cut values were chosen as a trade-off between high quality and large statistics. However, in the study of the systematic uncertainties, these cuts were relaxed to properly account for their effect.
Traces passing this selection are indicated by the grey dotted lines in Fig.~\ref{fig:clouds_fit}-bottom.

\begin{figure}
\includegraphics[width=\columnwidth]{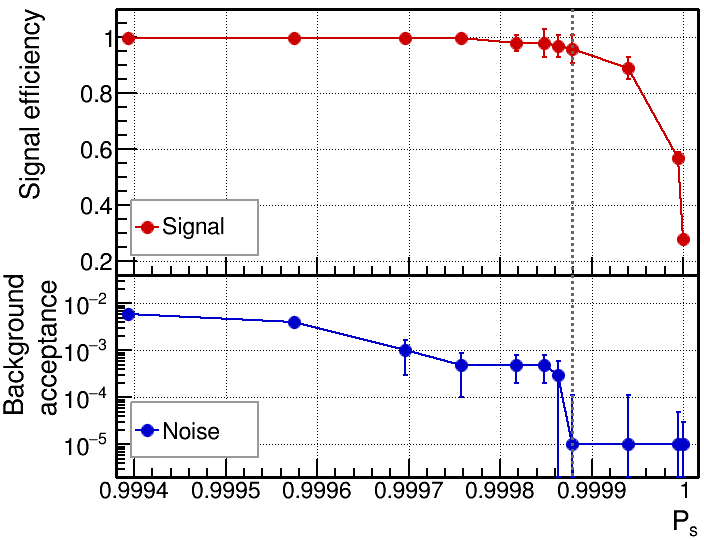}
\caption{Signal efficiency $\epsilon_{signal}$ and noise rejection efficiency $\epsilon_{noise}$ as a function of the signal probability $P_S$. The dotted line indicates the value selected for the analysis.} 
\label{fig:scanP}
\end{figure}

\subsection{Definition of $P_S$}
\label{sec:Psignal}
To discriminate radioactive events from the noise due to qubit decay, we further refine the data selection by computing a variable called \emph{ signal probability}, 
$P_S(k_{\mathrm{signal}}|N_{\mathrm{signal}},k_{\mathrm{control}},N_{\mathrm{control}})$.

In the following, $k_{\mathrm{signal}}$ and $k_{\mathrm{control}}$ denote the number of zeros observed in the signal and control windows, respectively, while $N_{\mathrm{signal}}$ and $N_{\mathrm{control}}$ denote their corresponding lengths.
In absence of radiation interaction, we assume a binomial model for the occurrence of zeros with an unknown probability $p$.

We first derive the posterior distribution of $p$ from the control region, $P(p|k_{\mathrm{control}},N_{\mathrm{control}})$. 
We then marginalize over $p$ the probability of observing $k_{signal}$ given the binomial model, namely $P(k_{signal}|N_{signal},p)$. To do so, we weight it by the posterior probability of $p$ as inferred from the control region, and integrate over $p$ itself. 
The left-tail probability is achieved by summing over all outcomes with a number of zeros lower than or equal to the observed value $k_{\mathrm{signal}}$. 
\begin{equation*}
\begin{split}
P_S(k_{\mathrm{signal}}|N_{\mathrm{signal}},k_{\mathrm{control}},N_{\mathrm{control}})
=\\
\sum_{k=0}^{k_{\mathrm{signal}}} \int_0^1
P(k|N_{\mathrm{signal}},p) 
P(p|k_{\mathrm{control}},N_{\mathrm{control}})
dp .
\end{split}
\end{equation*}

Large values of
$P_S(k_{\mathrm{signal}}|N_{\mathrm{signal}},k_{\mathrm{control}},N_{\mathrm{control}})$
therefore indicate observations that are unlikely to arise from background fluctuations alone.

\subsection{Optimization of $P_S(k_{\mathrm{signal}}|N_{\mathrm{signal}},k_{\mathrm{control}},N_{\mathrm{control}})$}

\label{sec:Pscan}
To determine the optimal value of $P_S(k_{\mathrm{signal}}|N_{\mathrm{signal}},k_{\mathrm{control}},N_{\mathrm{control}})$, hereafter denoted simply as $P_S$, we adopt the following procedure.

For each value of $P_S$, we construct the event histogram and perform a simultaneous fit to the fraction of events that pass the $P_S$ cut ($f_{\mathrm{passed}}$) and the fraction of rejected events ($f_{\mathrm{rejected}}$) to extract both the signal survival efficiency $\epsilon_{\mathrm{signal}}$ and the noise rejection efficiency $\epsilon_{\mathrm{noise}}$:

\begin{equation*}
f_{passed}  = \epsilon_{noise}N_{noise}\, B_{noise} + \epsilon_{signal}N_{signal}\, G_{signal}
\end{equation*}

and 

\begin{multline*}
    f_{rejected} =
    (1-\epsilon_{noise})N_{noise}\, B_{noise} \\
    + (1-\epsilon_{signal})N_{signal} G_{signal}
\end{multline*}
In this model, $B_{\mathrm{noise}}$ represents the binomial distribution used to describe the noise due to qubit decay, while $G_{\mathrm{signal}}$ is a Gaussian distribution used to model the expected signal.
To avoid potential bias arising from using the same dataset to optimize the P$_S$ cut and estimate the corresponding signal efficiency, these two steps were performed on statistically independent datasets, obtained by splitting the run into even- and odd-numbered triggered events.
Fig.~\ref{fig:scanP} shows the results of the scan on $P_S$, performed using the even events only.
This scan allowed to select the value of $P_S$ optimizing the signal-to-noise ratio. 
The odd events of the run were then used to extract the signal efficiency with the same method after this data selection (reported in the main text).

\begin{figure}
\centering
\includegraphics[width=\columnwidth]{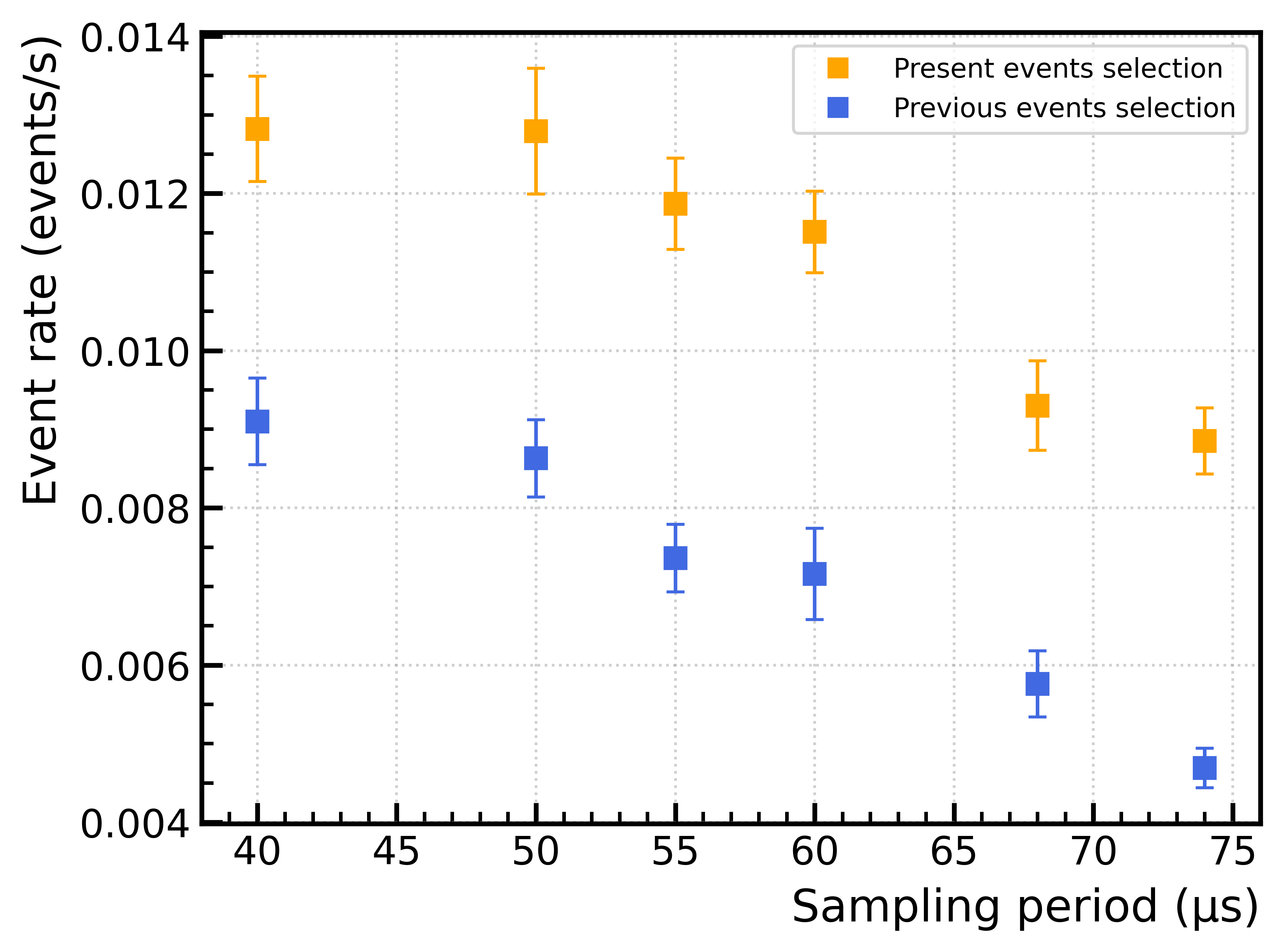}
\caption{Event rate measured at FNAL as a function of the sampling period. The expected event rate is (42$\pm$3)$\times$10$^{-3}$\,events/s.
Blue squares: rates obtained with the event selection technique described in~\cite{DeDominicis2026}. Orange squares: same rates computed with the analysis technique based on qubit state populations and $P_s$.  }
\label{fig:comparison_event_rates}
\end{figure}

In order to enable a more quantitative comparison between the previous and present techniques, we reanalyzed the data acquired in 2025 at the ``Quantum Garage" of the SQMS center at Fermilab (FNAL) and published in Ref.~\cite{DeDominicis2026}. 
The datasets were recorded using sampling periods of 40 $\mu$s, 50 $\mu$s, 55 $\mu$s, 60 $\mu$s, 68 $\mu$s, and 74 $\mu$s. 
At FNAL, the measurements were performed without a dedicated radiation source; consequently, the detected events originated primarily from environmental $\gamma$-rays and muons. 

Fig.~\ref{fig:comparison_event_rates} shows the average event rate as a function of the sampling period.
For each sampling period, the present analysis yields a consistently higher number of detected radiation-induced events compared to the previous technique, demonstrating a clear improvement in sensitivity. 
The improvement is particularly pronounced at longer sampling periods, where radiation-induced events are not well separated from background fluctuations (Fig.~5 of Ref.~\cite{DeDominicis2026}).
For shorter sampling intervals, where signal and noise are already well separated, the present analysis yields a $42\%$--$48\%$ improvement relative to the previous technique.

\subsection{Event Identification and Qubit Combination}
\label{sec:combination}
Figure~\ref{fig:histo_zero} shows the relaxations recorded by a single qubit in the 40-point signal region. Each histogram corresponds to a different day of measurement. In each spectrum, two distinct populations can be observed. 

The first, centered at approximately 7--9 zeros, is  associated with intrinsic qubit relaxation processes. Given the context of this work, this distribution is considered background or noise.
The second, smaller and centered at approximately 16--20 zeros, can be attributed to interactions of ionizing radiation originating from the $^{224}$Ra source. Indeed, such an event generates quasiparticle bursts whose influence persists for more than 1\,millisecond, manifesting as a larger number of relaxations. Moreover, its rate is of the same order of magnitude as that expected from $^{224}$Ra source, and its gradual decrease with time is consistent with the source decay constant ($\tau$=5.24\,days).
The candidate events appearing in the histograms of Figure~\ref{fig:histo_zero} were then selected following the approach described in the previous section.

\begin{figure}[t]
\includegraphics[width=\columnwidth]{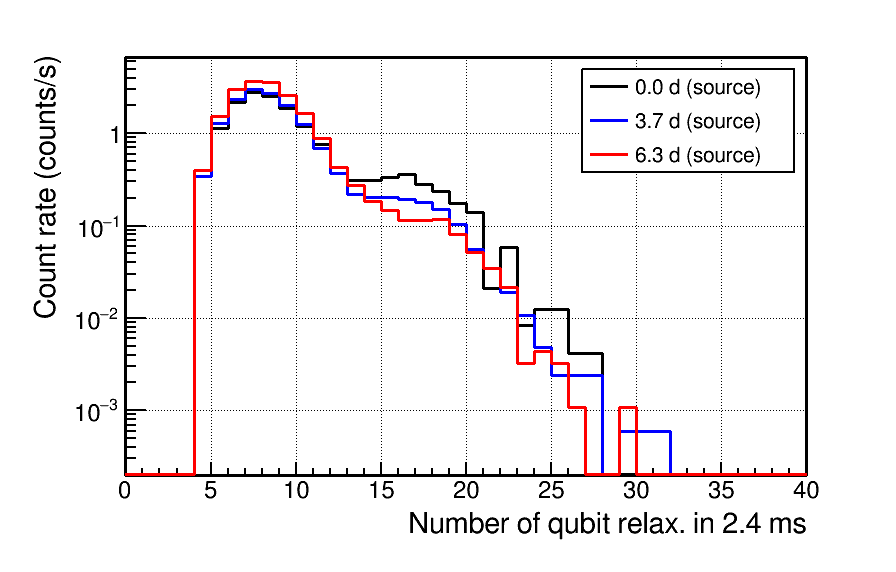}
\caption{Histogram of the number of qubit relaxations in 2.4\,ms observed within the signal region. The peak centered around 7--9 error counts is attributed to noise events arising from spontaneous qubit relaxation. The excess at higher error counts is produced by the radioactive sources placed near the chip ($^{224}$Ra) and environmental gamma radiation, with a count rate that decreases over time consistently with that of the source.} 
\label{fig:histo_zero}
\end{figure}

To enhance the detection efficiency, we used the more sensitive qubit (Q2) as the trigger and then we combined its signal with that measured by Q1. A joint output can increase the signal, if observed by both qubits, while suppressing the uncorrelated noise. This approach can better separate signal from the background, recovering weak signals (low number of zeros), immersed in the noise distribution. 
To this aim, for each trigger identified in Q2, we forced the acquisition of corresponding signal/control regions for Q1 at the same UnixTime. The number of zeros measured in the two signal windows of Q1 and Q2 was then summed to form a new signal. The result is shown in the gray histogram of Fig.~\ref{fig:qubit_combination}.
As before, we can identify two populations: a peak around 11--12 zeros, which corresponds to the noise measured by Q1+Q2, and a bump around 30--40 zeros, which corresponds to the summed signal.

\begin{figure}[t]
\includegraphics[width=\columnwidth]{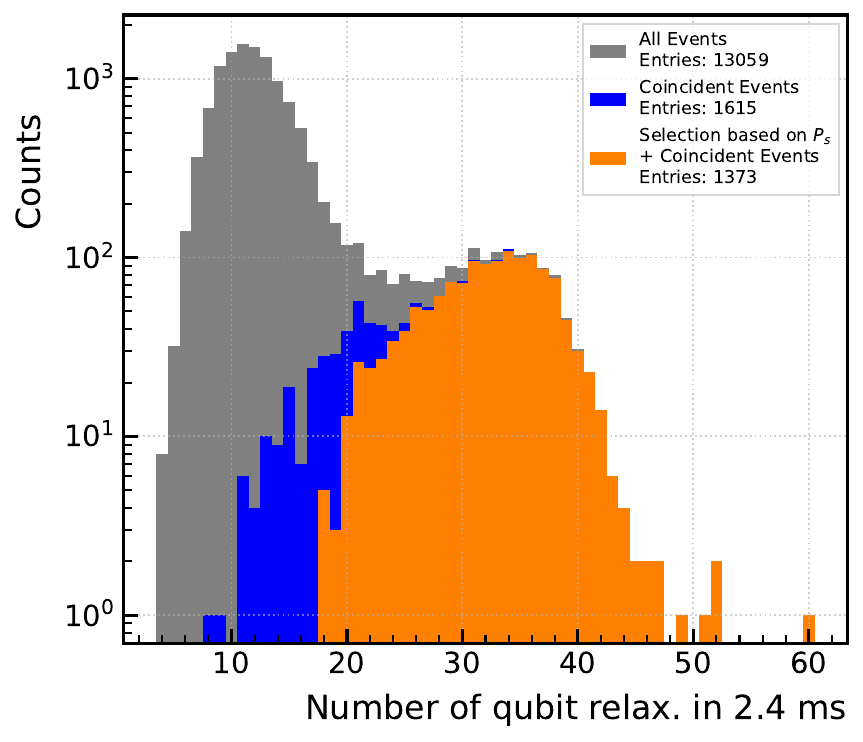}
\caption{Sum of the data stream recorded by Q1 and Q2 when Q2 triggers an event. Blue: only events in which the triggers Q1/Q2 are close to each other are retained. Orange: the data selection on P$_S$ is applied.} 
\label{fig:qubit_combination}
\end{figure}

At this point we selected events seen by both the qubits. Indeed, if only one of the two qubits detected a signal at a given UnixTime, the resulting summed output can be worsened with respect to the single qubit, since more noise is summed to the same signal.  
First, we investigated possible time delays between the signals detected by the two qubits. We searched for events triggered by Q1 within a 2.4-ms window around the Q2 trigger time, corresponding to the duration of the signal region. The procedure was repeated for every Q2 triggered event.
This study resulted in a distribution of the time difference between the Q1 and Q2 triggers centered at zero, showing no significant time delay between the two qubit responses. Based on this distribution, we defined a coincidence window around zero and required the Q1 trigger to occur within $\pm15$ samples from the Q2 trigger time.
Events lacking this condition were rejected, as the signal was detected by Q2 but not Q1. The chosen time window was meant to suppress random coincidences. The distribution of events obtained by imposing the time coincidence between the two qubits is shown in blue in Fig.~\ref{fig:qubit_combination}.

Finally, we repeated the radiation event selection procedure based on $P_{s}$ for the combined signals (orange in Fig.~\ref{fig:qubit_combination}) and reported the obtained rate as a function of time (green dots in Fig.~\ref{fig:decay}). 





\endgroup

\end{document}